\documentclass[5p]{elsarticle}

\usepackage{graphicx}
\usepackage{hyperref}
\usepackage{siunitx}
\newcommand{\SIhyp}[2]{\SI[number-unit-product={\text{-}}]{#1}{#2}}
\DeclareSIUnit\clight{\text{\ensuremath{c}}}
\biboptions{sort&compress}
\journal{Physics Letters B}

\begin{document}

\begin{frontmatter}

\title{Coexisting normal and intruder configurations in $^{32}$Mg}

\author[cns]{N.~Kitamura\corref{cor1}\fnref{fn1}}
\ead{nkitamur@utk.edu}
\cortext[cor1]{Corresponding author}
\fntext[fn1]{Present address: Department of Physics and Astronomy, University of Tennessee, Knoxville, Tennessee 37996, USA}
\author[csic,utokyo,cmu,nscl]{K.~Wimmer}
\author[uam]{A.~Poves}
\author[cns]{N.~Shimizu}
\author[surrey]{J.~A.~Tostevin}
\author[nscl,msu]{V.~M.~Bader}
\author[cmu]{C.~Bancroft}
\author[cmu]{D.~Barofsky}
\author[nscl,msu]{T.~Baugher}
\author[nscl]{D.~Bazin}
\author[nscl]{J.~S.~Berryman}
\author[guelph]{V.~Bildstein}
\author[nscl,msu]{A.~Gade}
\author[cns]{N.~Imai}
\author[tudarmstadt]{T.~Kr\"oll}
\author[nscl]{C.~Langer}
\author[cmu]{J.~Lloyd}
\author[nscl,msu]{E.~Lunderberg}
\author[iphc]{F.~Nowacki}
\author[cmu,nscl]{G.~Perdikakis}
\author[nscl]{F.~Recchia}
\author[cmu]{T.~Redpath}
\author[cmu]{S.~Saenz}
\author[nscl]{D.~Smalley}
\author[nscl,msu]{S.~R.~Stroberg}
\author[jaea,cns]{Y.~Utsuno}
\author[nscl]{D.~Weisshaar}
\author[cmu]{A.~Westerberg}

\address[cns]{Center for Nuclear Study, University of Tokyo, Wako, Saitama 351-0198, Japan}
\address[csic]{Instituto de Estructura de la Materia, CSIC, 28006 Madrid, Spain}
\address[utokyo]{Department of Physics, University of Tokyo, Bunkyo, Tokyo 113-0033, Japan}
\address[cmu]{Department of Physics, Central Michigan University, Mt.\ Pleasant, Michigan 48859, USA}
\address[nscl]{National Superconducting Cyclotron Laboratory, Michigan State University, East Lansing, Michigan 48824, USA}
\address[uam]{Departamento de F\'isica Te\'orica, Universidad Aut\'onoma de Madrid, 28049 Madrid, Spain}
\address[msu]{Department of Physics and Astronomy, Michigan State University, East Lansing, Michigan 48824, USA}
\address[surrey]{Department of Physics, University of Surrey, Guildford, Surrey GU2 7XH, United Kingdom}
\address[guelph]{Department of Physics, University of Guelph, Guelph, Ontario N1G 2W1, Canada}
\address[tudarmstadt]{Institut f\"ur Kernphysik, Technische Universit\"at Darmstadt, 64289 Darmstadt, Germany}
\address[iphc]{Institut Pluridisciplinaire Hubert Curien, 67037 Strasbourg, France}
\address[jaea]{Advanced Science Research Center, Japan Atomic Energy Agency, Tokai, Ibaraki 319-1195, Japan}

\begin{abstract}
Situated in the so-called ``island of inversion,'' the nucleus $^{32}$Mg is considered as an archetypal example of the disappearance of magicity at $N=20$. We report on high statistics in-beam spectroscopy of $^{32}$Mg with a unique approach, in that two direct reaction probes with different sensitivities to the underlying nuclear structure are employed at the same time. More specifically, states in $^{32}$Mg were populated by knockout reactions starting from $^{33}$Mg and $^{34}$Si, lying inside and outside the island of inversion, respectively. The momentum distributions of the reaction residues and the cross sections leading to the individual final states were confronted with eikonal-based reaction calculations, yielding a significantly updated level scheme for $^{32}$Mg and spin-parity assignments. By fully exploiting observables obtained in this measurement, a variety of structures coexisting in $^{32}$Mg was unraveled. Comparisons with theoretical predictions based on shell-model overlaps allowed for clear discrimination between different structural models, revealing that the complete theoretical description of this key nucleus is yet to be achieved.
\end{abstract}


\end{frontmatter}

While shell structure in atomic nuclei is the foundation of nuclear physics, the breakdown of the traditional magic numbers observed in unstable nuclei has revealed modifications in the shell structure and the important roles played by underlying nuclear interactions~\cite{SOR08,OTS01,OTS05,OTS10,OTS20}. In the ``island of inversion'' in the nuclear chart, ground states of neutron-rich $sd$-shell nuclei exhibit strong admixtures of intruder configurations involving the $fp$-shell orbitals, resulting in the disappearance of the canonical $N=20$ magic number~\cite{THI75,POV87,WAR90}. This concept was extended to other regions in the nuclear chart. To date, islands of inversion have been proposed at $N=8$, 20, 28, 40, and 50~\cite{SOR08,CAU14,LEN10,NOW16,NOW21}, and universal understanding of these islands and their driving mechanism has been one of the main topics in modern nuclear physics. This Letter focuses on the nucleus $^{32}$Mg ($N=20$ and $Z=12$), at the heart of the original island of inversion, which has received significant attention over the years.

The nucleus $^{32}$Mg has been studied by various approaches, e.g., $\beta$ decay of $^{32}$Na, Coulomb excitation, fragmentation reactions, and inclusive one-neutron knockout from $^{33}$Mg~\cite{DET79,NUM01,TRI08a,MOT95,CRA16,KAN10}. Most importantly, following the observation of the low excitation energy of the first $2^+$ state back in 1979~\cite{DET79}, the large quadrupole collectivity provided evidence for the strong deformation of the ground state and the quenched $N=20$ gap~\cite{MOT95}. In 2010, the first-excited $0^+$ state was identified in the $t$($^{30}$Mg,$p$) reaction~\cite{WIM10}, and later the existence of this state was confirmed in an in-beam $\gamma$-ray spectroscopy experiment~\cite{ELD19}. From a simplified point of view, the $0_2^+$ was interpreted as a counterpart of the near-spherical ground states of $^{30}$Mg~\cite{SCH09} and $^{34}$Si~\cite{ROT12}, and it was taken as a manifestation of shape coexistence, which is proposed to be the universal phenomenon in the islands of inversion~\cite{HEY11}. However, it was later shown that its structure is much more complex. Key properties of the structural change between the $0^+$ states are captured by the competition of different particle-hole configurations ($n$p$n$h). While low-lying states seem to be dominated by 0p0h and 2p2h configurations~\cite{SCH09,WIM10,ROT12}, theoretical studies pointed out the important role of 4p4h configurations~\cite{CAU14,MAC16}. In order to pin down the microscopic picture of the island of inversion, detailed spectroscopic information is essential.

Here, we report on in-beam $\gamma$-ray spectroscopy of $^{32}$Mg produced using two different direct reactions, $^9$Be($^{33}$Mg,$^{32}$Mg)$X$ one-neutron and $^9$Be($^{34}$Si,$^{32}$Mg)$X$ two-proton knockout, whose use as a spectroscopic tool has been established~\cite{GAD08b}. The advantage of this unique approach is that these two reactions are expected to have different selectivity for final states in $^{32}$Mg, because $^{34}$Si is considered to be of doubly-magic nature~\cite{BAU89,ROT12,BUR14,JON20}, whereas $^{33}$Mg is itself located inside the island of inversion where particle-hole configurations dominate in the ground states~\cite{YOR07}. In the present Letter, we discuss the main results of the experiment and implications for the understanding of the structure of nuclei in the island of inversion. For more details on the experimental results and theoretical calculations, the reader is referred to Ref.~\cite{KIT21}.

\begin{figure*}[t!]
\includegraphics[scale=0.5]{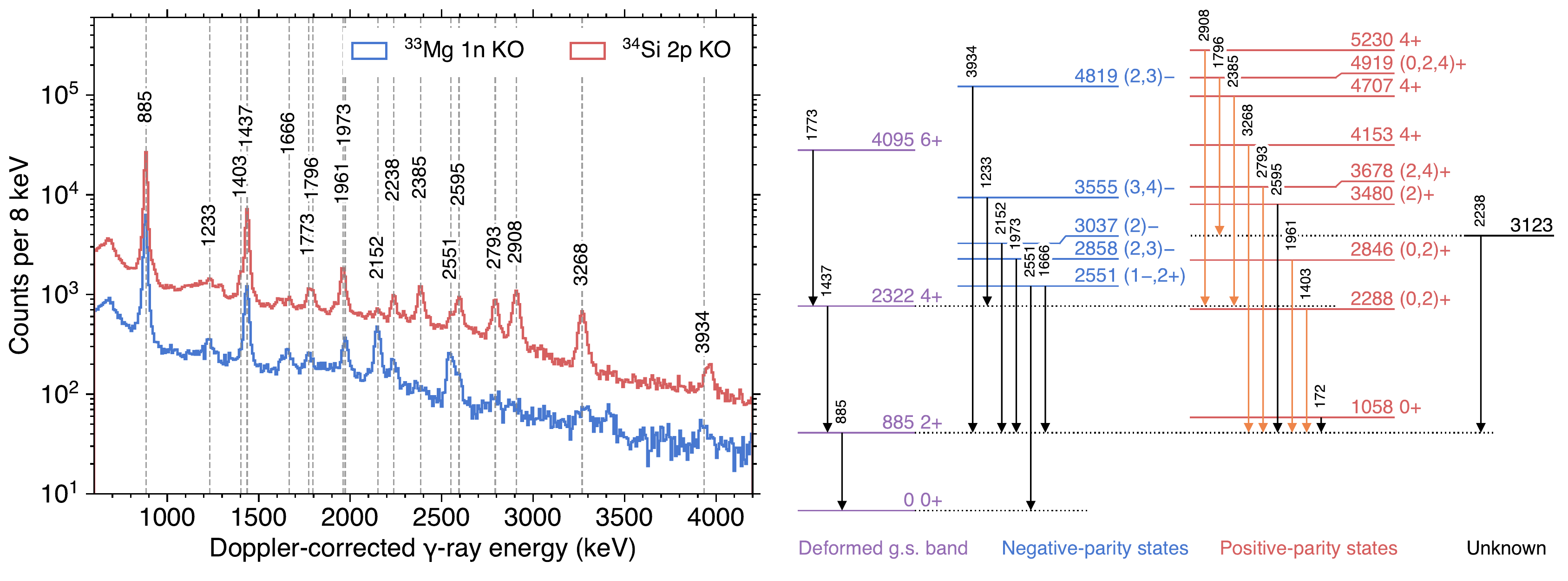}
\caption{(a) Doppler-corrected add-back $\gamma$-ray spectra for $^{32}$Mg recorded in coincidence with incoming $^{33}$Mg (blue) and $^{34}$Si (red) projectiles. The transitions placed in the level scheme are indicated by vertical dashed lines and labeled by their energies. (b) Level scheme of $^{32}$Mg constructed in the present work. Assigned spin-parities are shown besides the level energies. Newly established transitions are indicated by the orange arrows. The level scheme has been constructed using $\gamma$-$\gamma$ coincidences and guidance from previous studies (see text for details).}
\label{fig:gammaandlevels}
\end{figure*}

The experiment was performed at the National Superconducting Cyclotron Laboratory at Michigan State University. A $^{48}$Ca beam at \SI{140}{MeV/nucleon}, delivered from the Coupled Cyclotron Facility, impinged on a \SIhyp{846}{mg/cm^2} thick $^9$Be target to produce secondary beams of $^{33}$Mg and $^{34}$Si. The purification of the projectile-like particles was achieved by a \SIhyp{300}{mg/cm^2} thick aluminum wedge degrader in the A1900 fragment separator~\cite{MOR03}. The beams were then transported to the S800 spectrograph~\cite{BAZ03a} and focused on a \SIhyp{375}{mg/cm^2} thick $^9$Be secondary target to induce the one-neutron (two-proton) knockout reactions from $^{33}$Mg ($^{34}$Si). The beams had average on-target intensities of \num{620} and \SI{5.2e5}{s^{-1}}, purities of \num{15} and \SI{66}{\percent}, and incident energies of \num{99.6} and \SI{94.8}{MeV/nucleon}, respectively. The particle identification for the incoming and outgoing ions was performed on an event-by-event basis using the standard A1900 and S800 focal-plane detectors~\cite{YUR99}. The inclusive cross sections were measured to be $\sigma_{1n}^\mathrm{incl} = \SI{104(5)}{mb}$ and $\sigma_{2p}^\mathrm{incl} = \SI{0.96(8)}{mb}$. These results agree with previous measurements of the same reaction~\cite{BAZ03b,ZWA05} and in this mass region~\cite{TER08,FER18,KIT20}. A set of parallel-plate avalanche counters installed at the dispersive plane, just upstream of the reaction target, was used to determine the momentum of the incoming ions, which is beneficial to achieve a higher resolution for the momentum distributions of the reaction residues. The reaction target was surrounded by the state-of-the-art array of $\gamma$-ray detectors, Gamma-Ray Energy Tracking In-beam Array, GRETINA~\cite{LEE04,FAL16,WEI17}. At the time of the experiment, the array consisted of seven modules, each housing four high-purity germanium crystals and it was set up with four modules at \SI{58}{\degree} and the remaining three at \SI{90}{\degree} with respect to the beam axis. The large solid-angle coverage of the array gave a photo-peak efficiency of about \SI{5}{\percent} at \SI{1}{MeV}. To obtain the absolute $\gamma$-ray intensities from experimental spectra, response functions generated by a dedicated Monte Carlo package~\cite{RIL21} were used. More details on the experimental setup have been given elsewhere~\cite{KIT20,KIT21}.

Doppler-corrected $\gamma$-ray spectra obtained from the two reactions are shown in Fig.~\ref{fig:gammaandlevels}. The nearest-neighbor add-back procedure was employed to improve the peak-to-total ratio, which is beneficial for the identification of individual peaks. As evident in Fig.~\ref{fig:gammaandlevels}, the two reactions present very different population of states. Several new $\gamma$ lines were identified in the two-proton knockout ($2p$ KO) spectrum, while the majority of the peaks seen in the one-neutron knockout ($1n$ KO) spectrum correspond to those reported in $\beta$-decay measurements~\cite{MAT07,TRI08a}.

Following the $\gamma$-$\gamma$ coincidence analysis, the level scheme for $^{32}$Mg was significantly updated as shown in Fig.~\ref{fig:gammaandlevels}. For weak transitions, e.g., the \num{1666}- and \SIhyp{3934}{keV} $\gamma$ rays, the results from previous experiments~\cite{MAT07,TRI08a} were utilized to guide the placements. We emphasize that the high $\gamma$-ray resolving power enabled by GRETINA was particularly beneficial for the placements of the \num{1403}- and \SIhyp{1961}{keV} transitions seen in the two-proton knockout spectrum. The \SIhyp{1403}{keV} transition, lying close to the strong \SIhyp{1437}{keV} $4^+_1\to 2^+_1$ peak, was found to be in coincidence with the \SIhyp{885}{keV} transition, and this observation identifies a new state at \SI{2288}{keV}, situated slightly below the $4^+_1$ state. Similarly, the \SIhyp{1961}{keV} transition turned out to be different when using the peak at \SI{1973}{keV} observed in the one-neutron knockout channel, and following the $\gamma$-$\gamma$ result, this transition was found to feed the $2^+_1$ state, establishing a new state at \SI{2846}{keV}.

\begin{figure}[t!]
\includegraphics[width=\columnwidth]{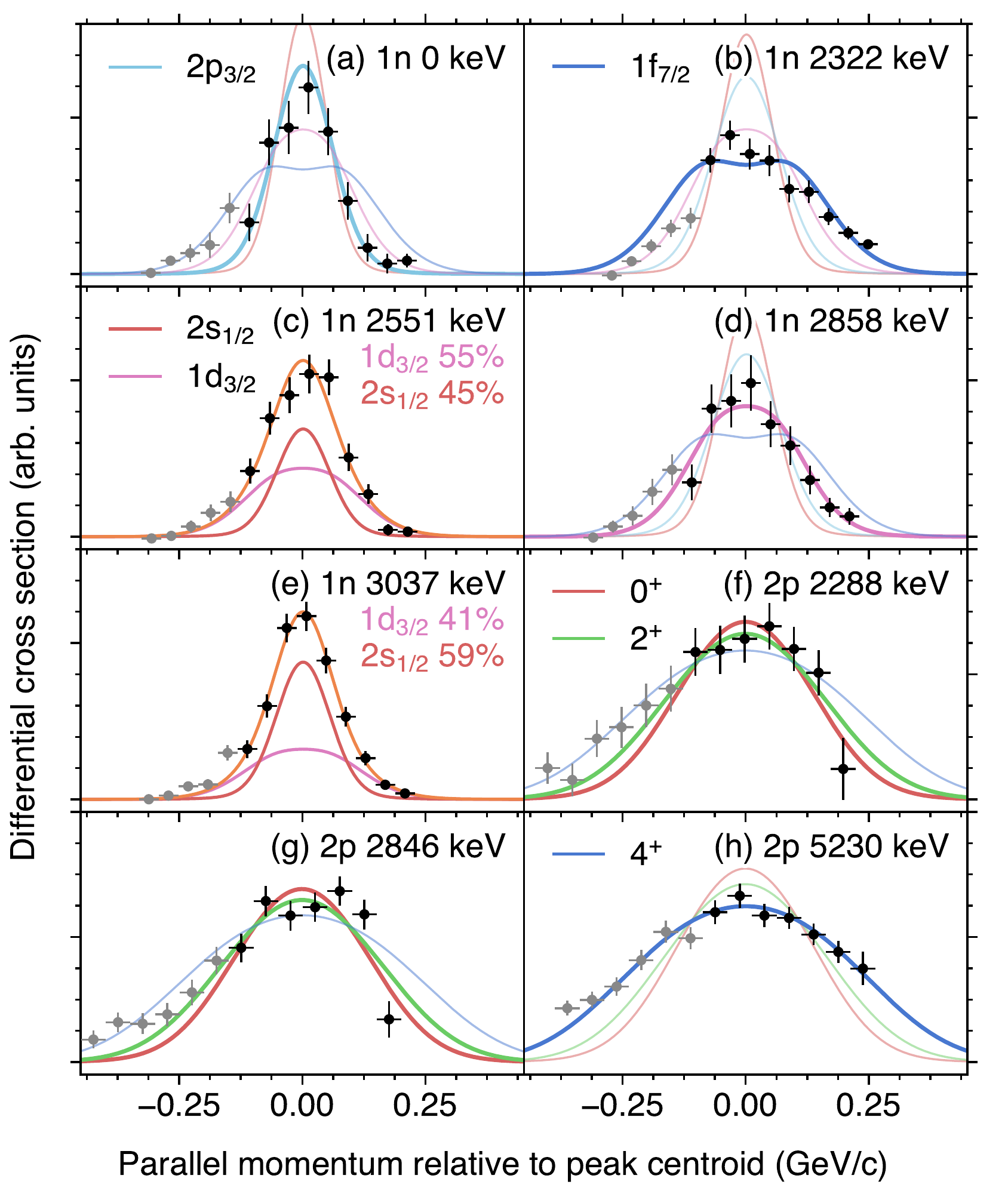}
\caption{(a--e) Selected momentum distributions for the one-neutron knockout reaction from $^{33}$Mg. For the states at (c) \num{2551} and (e) \SI{3037}{keV}, the experimental distributions are fitted by a combination of two different orbitals, and the orange curve represents the sum of these two contributions. Note that only the $(1d_{3/2},2s_{1/2})$ combination corresponding to the $1^-$ spin-parity assignment is shown for the \SIhyp{2551}{keV} state. (f--h) Selected momentum distributions for the two-proton knockout reaction from $^{34}$Si. Since the low-momentum side of the distribution is influenced by indirect processes that are not included in the reaction modeling, the gray data points have been ignored in the determination of the angular momentum transfer. The theoretical curves have been normalized to the area under the black data points.}
\label{fig:momdist}
\end{figure}

\begin{figure}[t!]
\includegraphics[scale=0.5]{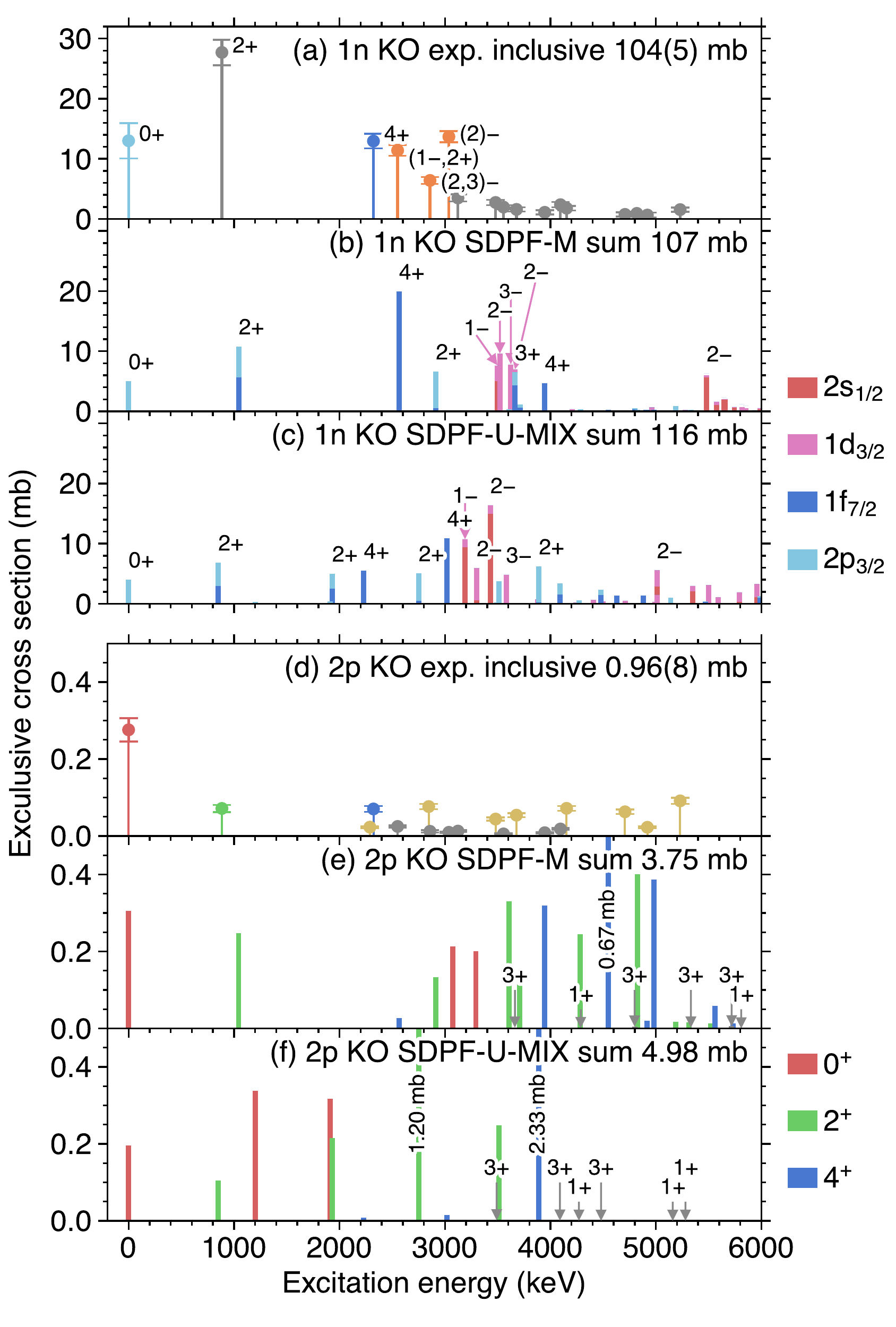}
\caption{Experimental (a--c) one-neutron and (d--f) two-proton knockout cross sections in comparison with theoretical calculations. (a--c) The colors represent contributions of different single-particle orbitals. Some states, shown in orange in panel (a), are populated by a combination of different angular momenta, and the compositions of different orbitals are shown in Fig.~\ref{fig:momdist}. (d--f) The bars are colored according to their spin-parity. The yellow bars in panel (d) represent states strongly populated in two-proton knockout, thus candidates for either $0^+$, $2^+$, or $4^+$. Note that the cross section populating the $0^+_2$ state is always included the ground-state cross section, because of the long lifetime of this state. In panels (a, d), states that are not used for the main discussion in this Letter are shown in gray.}
\label{fig:xsec}
\end{figure}

The one-neutron knockout reaction possesses sensitivity to the single-particle orbital from which the neutron is removed, and therefore, it serves as a powerful tool to assign spins and parities. We note that the one-neutron knockout reaction can produce final states in $^{32}$Mg with contributions from different orbitals due to the coupling of the removed nucleon with  the finite spin of the $^{33}$Mg ground state, unlike in one-neutron knockout from the $J^\pi=0^+$ ground state of even-even nuclei where the transferred orbital angular momentum ($\Delta L$) can uniquely be determined from the momentum distribution. The experimental momentum distributions were extracted (see Fig.~\ref{fig:momdist}), taking into account the contributions from feeding states with proper weighting factors, as is done when extracting final-state exclusive cross sections (details can be found, e.g., in Ref.~\cite{KIT20}). The experimental distributions are then compared with theoretical calculations based on the eikonal reaction theory~\cite{HAN03,GAD08a} folded with the experimental resolution (\SI{0.08}{GeV/\clight} FWHM). This procedure allows us to investigate the ground-state spin-parity of $^{33}$Mg. While a magnetic moment measurement indicated a $J^\pi=3/2^-$ assignment for the $^{33}$Mg ground state~\cite{YOR07}, $\beta$-decay studies suggested $J^\pi=3/2^+$~\cite{NUM01,TRI08b}. This disagreement has been a controversial topic as the spin-parity assignment serves as a first test of the underlying configuration~\cite{NEY11}. Recently, $J^\pi=3/2^-$ was reported in a one-neutron knockout measurement from $^{34}$Mg~\cite{BAZ21}, and the debate is now converging. The experimental ground-state momentum distribution extracted from the present data is best compatible with the calculation assuming one-neutron knockout from the $2p_{3/2}$ orbital, as can be seen in Fig.~\ref{fig:momdist}(a), supporting the $3/2^-$ assignment, in agreement with Refs.~\cite{YOR07,KAN10,BAZ21}. Also, the momentum distribution extracted for the $4^+_1$ state at \SI{2322}{keV}, shown in Fig.~\ref{fig:momdist}(b), presents a broad structure and is only compatible with $1f_{7/2}$, as expected from the spin-coupling to yield a spin-parity of $4^+$.

Next, we present new insight into the wave function composition of the $^{33}$Mg ground state. In Ref.~\cite{YOR07}, the spin-parity of $3/2^-$ was discussed in terms of the manifestation of a 3p2h configuration. One-neutron knockout cross sections populating specific final states measured in the present experiment allow us to revisit this discussion. The experimental exclusive cross sections are plotted in Fig.~\ref{fig:xsec}. Two noteworthy features are the large cross sections for the ground state and the $4^+_1$ state. These two states are exclusively populated by knockout from the $2p_{3/2}$ and $1f_{7/2}$ orbitals, respectively, and their large cross sections directly imply a large neutron occupation of both these orbitals in the ground state of $^{33}$Mg. Combined with the $3/2^-$ ground-state assignment, this result clearly demonstrates that the 3p2h configuration dominates the $^{33}$Mg ground state. We emphasize that the 3p2h configuration is seen as a particle in the neutron $fp$ shell coupled to the 2p2h intruder configuration. Of particular interest is the substantial $2p_{3/2}$ occupation in the $^{33}$Mg ground state, which was also found to be large in the ground state of $^{32}$Mg~\cite{TER08}. This is a natural consequence of the narrowing of the gap between the $2p_{3/2}$ and $1f_{7/2}$ orbitals~\cite{CAT10,CAU14} and thus the quenching of the canonical $N=20$ and $N=28$ shell closures, and it provides further insight into the microscopic origin of the island of inversion.

Following the establishment of the 3p2h nature of the ground state of $^{33}$Mg, the one-neutron knockout cross sections can be used as indicators for other intruder states in $^{32}$Mg. Because the removal of a neutron from the $sd$ orbitals in $^{33}$Mg leads to negative-parity final states, the observed large cross sections populating the \num{2551}-, \num{2858}-, and \SIhyp{3037}{keV} states are a signature of 3p3h intruder-like negative-parity levels~\cite{TRI08a,CAU14}.

For the \SIhyp{3037}{keV} state, as can be seen in Fig.~\ref{fig:momdist}(e), the narrow momentum distribution showing a large $2s_{1/2}$, i.e., $\Delta L=0$ contribution and the non-observation of a ground-state decay branch suggests a $(2)^-$ assignment. For the \SIhyp{2858}{keV} state, combining the indication from the $\beta$-decay work~\cite{TRI08a}, this state is also firmly identified as a negative-parity state and the $(2,3)^-$ assignment is adopted in this work. The momentum distribution shown in Fig.~\ref{fig:momdist}(d) is compatible with a single $1d_{3/2}$ removal calculation. A $1^-$ assignment is likely for the \SIhyp{2551}{keV} state, as this state decays both to the ground and $2^+_1$ states. However, a $2^+$ assignment to this state~\cite{TRI08a} cannot be ruled out based on the momentum distribution, shown in Fig.~\ref{fig:momdist}(c). In this specific case of a close-lying doublet, the gate for the \SIhyp{2551}{keV} peak has been placed on only the low-energy side of the peak. The impact of the width and location of the $\gamma$-ray gate on the momentum distribution has been carefully investigated, minimizing contamination.

The lowest firmly-assigned negative-parity state in $^{32}$Mg was established at \SI{2858}{keV} in the present work and this observation further extends the systematics. The excitation energy of the lowest negative-parity state drops when going from $^{34}$Si to $^{32}$Mg by \SI{1.4}{MeV}~\cite{LIC19}, and this observation is tied to the reduction of the effective size of the $N=20$ shell gap~\cite{KIT20}. It is worth noting that the excitation-energy drop at $^{32}$Mg is considered to be boosted by correlation effects~\cite{CAU14}, unlike the systematics established for the $N=18$ isotones~\cite{KIT20}.

\begin{figure}[t!]
\includegraphics[scale=0.5]{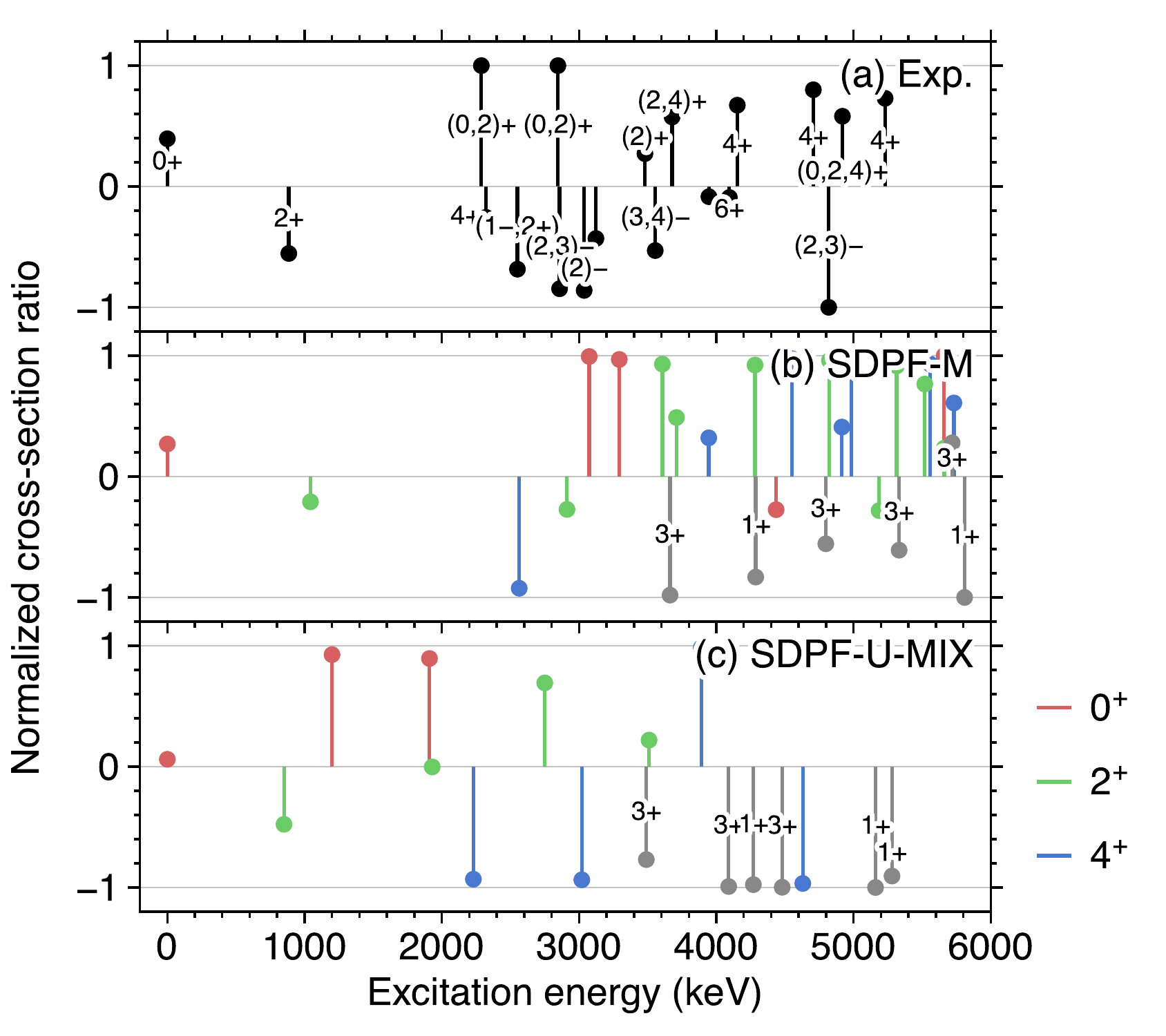}
\caption{Normalized ratios of (a) experimental one-neutron and two-proton knockout cross sections in comparison with those taken from eikonal-based reaction calculations coupled with shell-model overlaps from (b) SDPF-M and (c) SDPF-U-MIX.}
\label{fig:xsecratio}
\end{figure}

Going one step further, we demonstrate that the use of the two different reactions serves as a powerful tool to elucidate the states' nature. From Fig.~\ref{fig:gammaandlevels}, it is clearly seen that some of the transitions appear prominently or even exclusively in the spectrum of one of the reactions. To make this argument more quantitative, the experimental exclusive cross sections were analyzed following the approach of Ref.~\cite{ELM19}. In Fig.~\ref{fig:xsecratio}, normalized cross-section ratios, $(R_{2p}-R_{1n})/(R_{2p}+R_{1n})$, with relative exclusive cross sections defined as $R_{2p}=\sigma_{2p}(J^\pi_f)/\sigma_{2p}^\mathrm{incl}$ and $R_{1n}=\sigma_{1n}(J^\pi_f)/\sigma_{1n}^\mathrm{incl}$, are shown. Negative (positive) values of the ratio indicate strong population in the one-neutron (two-proton) knockout reaction.

The negative-parity states discussed earlier, the \num{2551}-, \num{2858}-, and \SIhyp{3037}{keV} states, are indeed shown to be characterized by large negative values, validating the use of the cross-section ratio. It can be seen in Fig.~\ref{fig:xsecratio} that the negative-parity candidates at \num{3555} and \SI{4819}{keV}~\cite{TRI08a} also have large negative values. Likewise, this analysis aids the spin-parity assignments for states with large positive values, i.e., those strongly populated in two-proton knockout. In this reaction, to first order, two protons are removed from the $1d_{5/2}$ orbital, and thus the final-state spin-parity is limited to $0^+$, $2^+$, or $4^+$.

While this method gives constraints in the spin-parity assignments in a model-independent manner, theoretical calculations can justify this approach. To demonstrate this, final-state exclusive cross sections for two-proton knockout are calculated by employing the reaction dynamics based on the eikonal approximation~\cite{TOS04,TOS06,SIM09a,SIM09b} that take into account two-nucleon amplitudes (TNAs) from shell-model calculations. Similarly, one-neutron knockout cross sections were calculated using the same framework with shell-model spectroscopic factors (SFs)~\cite{HAN03,GAD08a}. The SFs and TNAs were calculated in the shell model employing the SDPF-M~\cite{UTS99} and SDPF-U-MIX~\cite{CAU14} interactions. The former includes only the $1f_{7/2}$ and $2p_{3/2}$ orbitals for protons and neutrons in addition to the $sd$-shell orbitals, the latter has the full neutron $sd$- and $fp$-shell degrees of freedom and protons confined in the $sd$ shell, making it one of the state-of-the-art interactions. It also accurately reproduces the low excitation energy of the $0^+_2$ state and the low-lying level structure of the neighboring nuclei, i.e., $^{30}$Mg and $^{34}$Si, and is often employed for spectroscopic studies around this region (see, for example, Refs.~\cite{ROT12,FER18,LIC19}). Combining the shell model results with the reaction model calculations the exclusive cross sections were obtained, as shown in Fig.~\ref{fig:xsec}. The cross-section ratios, presented in Fig.~\ref{fig:xsecratio}, show large positive values for states having spin-parities of either $0^+$, $2^+$, or $4^+$, and large negative values for positive-parity odd-spin states.

The momentum distributions in two-proton knockout were analyzed~\cite{SAN11,LON20} to better constrain the spin-parity assignments for the $(0,2,4)^+$ states highlighted in the cross-section-ratio discussion. Theoretical momentum distributions calculated using the framework described above were folded with the experimental resolution, which includes a Heaviside function to account for the energy loss in the target. The overall resolution is \SI{0.26}{GeV/\clight} FWHM. Due to the smearing effect introduced by the energy loss in the thick target used in the present experiment, both the high and low momentum sides suffered from acceptance cutoffs in the S800. Nevertheless, the experimental distributions allow for the discrimination between $4^+$ and $(0,2)^+$ states. Based on comparisons with the theoretical momentum distributions, $4^+$ assignments were firmly made for the \num{4153}-, \num{4707}-, and \SIhyp{5230}{keV} states. $(0,2)^+$ assignments are proposed for the \num{2288}- and \SIhyp{2846}{keV} states, as shown in Figs.~\ref{fig:momdist}(f) and (g). Either of these could correspond to the $0^+_3$ state predicted in Ref.~\cite{MAC16}, based on their excitation energies and the non-observation of the ground-state decay branches.

To gain further insight into the theoretical description of $^{32}$Mg, the experimental levels and their one-neutron knockout cross sections are compared with shell-model calculations (see Fig.~\ref{fig:xsec}). While a quenching factor of \num{0.9} is expected for the one-neutron knockout reaction based on the systematics~\cite{TOS14,TOS21}, such a factor was not applied to the cross sections presented in Fig.~\ref{fig:xsec}. Although the predicted lowest negative-parity state is lying higher than the observation at \SI{2858}{keV}, the SDPF-M calculation reasonably reproduces the location of negative-parity states and also their cross sections, as was the case for $^{30}$Mg~\cite{KIT20}. The calculation shows that the 3p3h contributions of the lowest negative-parity states lie around \SI{90}{\percent}, highlighting their intruder nature. Additionally, the sizable $4^+_1$ cross section observed in one-neutron knockout shows good agreement with the prediction. The SDPF-U-MIX calculations present better level energy predictions, including the $0^+_2$ state, as compared to SDPF-M. Also, the calculated one-neutron knockout cross sections agree well with the observation. However, the difference seen in the two-proton knockout cross sections is striking. The SDPF-U-MIX prediction, shown in Fig.~\ref{fig:xsec}(f), exhibits much more concentrated strengths than the observation and the SDPF-M prediction. In SDPF-U-MIX, the large cross sections for the $2^+_3$ and $4^+_3$ are arising from substantial overlaps creating proton holes in the $1d_{5/2}$ orbital, as expected from a 0p0h component of \SI{90}{\percent} in the $^{34}$Si ground state, whereas it is diminished in SDPF-M down to \SI{60}{\percent}, presenting a less doubly-magic nature. The root of the disagreement remains to be understood. In addition, the fact that both calculations require a significant quenching factor of around \num{0.2}, as compared to the typical value of \num{0.5} found in normal $sd$-shell test cases~\cite{TOS06}, to reconcile the observed inclusive cross section questions the present theoretical description of both $^{34}$Si and $^{32}$Mg.

In this work, the $^{33}$Mg ground state was established as being dominated by an intruder-dominated 3p2h configuration where three neutrons in the $fp$ orbitals couple to $J^\pi=3/2^-$. More specifically, a large fraction of the underlying configurations were shown to be of the $(1f_{7/2})^2(2p_{3/2})^1$ neutron configuration, quantitatively supported by the experimental SFs and the shell-model calculations. For $^{32}$Mg, the observations can be understood in terms of the coexistence of a variety of structures, which fall into three categories (see also Fig.~\ref{fig:gammaandlevels}). (i) The $2^+_1$, $4^+_1$, and $6^+_1$ states built on top of the ground state are, as discussed in Ref.~\cite{CRA16}, interpreted as a strongly-deformed rotational band, which requires a sizable mixing of 2p2h contributions and possibly 4p4h ones~\cite{CAU14}. (ii) The negative-parity states populated in this work are interpreted as having predominant 3p3h configurations made by particle-hole excitations coupled to intruder 2p2h configurations. The comparison with the shell-model calculation reinforces the claim that the negative-parity states are described as intruder. (iii) A group of $0^+$, $2^+$, and $4^+$ states, strongly populated in the two-proton knockout reaction, is considered to be formed by excitations in the proton side while the neutrons remain confined to the $sd$ shell. Therefore, the microscopic structure is distinct from those of (i) and (ii). These states are candidates for the members of the normal-configuration bands which are embedded in the off-yrast region and are hitherto unexplored, and the present measurement provides an additional indication for the coexistence of normal configurations, presumably characterized by near-spherical shapes. In a simple interpretation, the $0^+_2$ state is also seen as a normal, spherical band head that coexists with the deformed ground-state band~\cite{WIM10}. The complete observation of the band structure and transitions between them will shed more light on the nature of these states and their mixing.

To conclude, we explored the states in $^{32}$Mg using two different direct reactions. From the one-neutron knockout reaction on $^{33}$Mg, the 3p2h nature and the spin-parity of $3/2^-$ for the $^{33}$Mg ground state were put on a firm footing. By combining the two-proton knockout data, valuable information on the structure of this key nucleus was gained. The level scheme for $^{32}$Mg and their spin-parity assignments have been updated, and as a result, a variety of structures coexisting in $^{32}$Mg was unraveled. The states can be classified as (i) the strongly-deformed ground-state band dominated by intruder configurations, (ii) 3p3h negative-parity intruder levels, and (iii) hitherto-unexplored high-lying $0^+$, $2^+$, and $4^+$ states characterized by normal configurations. Notably, this work presented the first indication for the $0^+_3$ state that was theoretically predicted. Additionally, the lowest firmly-assigned negative-parity state in $^{32}$Mg was established at \SI{2858}{keV}, and this observation helps us further extend the systematics in the $N=20$ isotones. It was found that the experimental level structure and cross sections do not dramatically deviate from the shell-model prediction by the SDPF-M interaction aside from the $0^+_2$ excitation energy. The observed two-proton knockout cross sections are very different from the SDPF-U-MIX prediction, even though this interaction can reproduce the $0^+_2$ excitation energy, posing further questions concerning the delineation of the island of inversion.

\section*{Acknowledgments}
We express our gratitude to the accelerator staff at NSCL for their efforts in beam delivery during the experiment. N.K.\ acknowledges support of the Grant-in-Aid for JSPS Fellows (18J12542) from the Ministry of Education, Culture, Sports, Science, and Technology (MEXT), Japan. K.W.\ acknowledges support from the Ministerio de Ciencia, Innovaci\'on y Universidades (Spain) RYC-2017-22007. A.P.\ is supported in part by the Ministerio de Ciencia, Innovaci\'on y Universidades (Spain), Severo Ochoa Programme SEV-2016-0597 and grant PGC-2018-94583. The SDPF-M calculations were enabled by the CNS-RIKEN joint project for large-scale nuclear structure calculations and were performed mainly on the Oakforest-PACS supercomputer. N.S.\ acknowledges support from ``Priority Issue on post-K computer'' (hp190160) and ``Program for Promoting Researches on the Supercomputer Fugaku'' (hp200130) by JICFuS and MEXT, Japan. J.A.T.\ acknowledges support from the U.K.\ Science and Technology Facilities Council Grant No.\ ST/L005743/1. This work was supported by the U.S.\ Department of Energy (DOE), Office of Science, Office of Nuclear Physics, under Grant No.\ DE-SC0020451 and by the U.S.\ National Science Foundation (NSF) under Grant No.\ PHY-1306297. GRETINA was funded by the U.S.\ DOE, Office of Science. Operation of the array at NSCL was supported by the U.S.\ NSF under Cooperative Agreement No.\ PHY-1102511 (NSCL) and DOE under Grant No.\ DE-AC02-05CH11231 (LBNL).

\end{document}